\documentclass[twocolumn,superscriptaddress,prb]{revtex4-2}

\usepackage{dcolumn}
\usepackage{bm}
\usepackage{subfigure}
\usepackage[normalem]{ulem}
\usepackage{enumerate}
\usepackage{cancel}
\usepackage{flushend}
\usepackage{bbold}
\usepackage{mathtools}
\usepackage{float}
\usepackage{stmaryrd}
\usepackage{colortbl}
\usepackage[table]{xcolor}
\usepackage{array,ragged2e}
\usepackage{wasysym,graphicx,multirow,textcomp}
\usepackage{url}
\usepackage{romannum}
\usepackage{tabularx}
\usepackage[final]{changes}
\usepackage[colorlinks = true,linkcolor = red,citecolor = magenta]{hyperref}
\usepackage[sort&compress]{natbib}

\newcommand{\figsizeone}{0.8}

\begin{document}

\draft

\title{Dynamics in non-Hermitian systems with nonreciprocal coupling}

\author{Jung-Wan Ryu}
\email{jungwanryu@gmail.com}
\address{Center for Theoretical Physics of Complex Systems, Institute for Basic Science (IBS), Daejeon 34126, Republic of Korea}

\date{\today}

\begin{abstract}
We reveal that non-Hermitian Hamiltonians with nonreciprocal coupling can achieve amplification of initial states without external gain due to a kind of inherent source. We discuss the source and its effect on time evolution in terms of complex eigenenergies and non-orthogonal eigenstates. Demonstrating two extreme cases of Hamiltonians, namely one having complex eigenenergies with orthogonal eigenstates and one having real eigenenergies with non-orthogonal eigenstates, we elucidate the differences between the amplifications from complex eigenenergies and from non-orthogonal eigenstates. 
\end{abstract}

\maketitle

\section{Introduction}

A nonreciprocal system is one in which the transmission of a state between any two positions depends on the direction of propagation, i.e., source and measure points are not interchangeable. In nonreciprocal systems, the off-diagonal scattering parameters of the s-matrix, which relates the initial and final states in source and measure points of a physical system, are not equivalent, or in other words $S_{12} \neq S_{21}$. Nonreciprocal systems have been realized in various fields of physics such as electromagnetics \cite{Xu19, Pot04, Est14, Rue16, Kod13}, acoustics \cite{Fle14, Wan18}, electronics \cite{Sou18}, quantum systems \cite{Pea16, Lau18}, and mechanics \cite{Nas15, Cou17, Bra19}. Recently, non-Hermitian topological models with nonreciprocal coupling terms have also attracted a lot of attention since nonreciprocity induces all the eigenstates to localize at the system boundary, a feature called the non-Hermitian skin effect \cite{Hat96, Hat97, Yao18a, Yao18b, Yan19, Jin19, Wan19, Lee19, Gha19, Gha21}.

Dynamics in non-Hermitian systems can be very different from that in Hermitian systems because of non-normality as well as complex eigenvalues. In non-normal systems with non-orthogonal eigenstates, one of the most interesting behaviors is that initial states amplify with time in the transient regime, which is called transient growth \cite{Tre05}, and finally amplify or decay exponentially with time in the long-time regime. Such transient growth and non-normality have been previously considered in different fields, e.g., fluid mechanics \cite{Tre93, Tre97}, spatial pattern formations \cite{Neu02, Rid11}, stochastic dynamics \cite{Bia17}, photonic media \cite{Mak14, Mak21}, and complex networks \cite{Asl18a, Asl18b, Bag20, Mou21, Dua22}.

In this work, we study time evolution in nonreciprocal systems. We discuss the inherent source in nonreciprocal systems by regarding them as coupled systems with unidirectional coupling, i.e., master and slave systems. The inherent source creates amplifications of initial states in time that are associated with complex eigenvalues and non-orthogonal eigenvectors in nonreciprocal systems without external gain. To elucidate the temporal behaviors, we demonstrate various time evolutions of specific initial states in nonreciprocal systems with unidirectional coupling under different boundary conditions. Finally, we reveal that non-Hermitian Hamiltonians with nonreciprocal coupling can achieve amplification of initial states without external gain due to the inherent source.

The paper is structured as follows. In Section II, we study the reason for the existence of an inherent source in nonreciprocal systems and introduce models with unidirectional coupling under a periodic boundary condition (PBC) and open boundary condition (OBC). Section III shows the results of time evolutions in the models. We demonstrate transient growth and long-time exponential behavior using different measures defined by right eigenstates, left eigenstates, and biorthogonal states. In Section IV, we summarize our results.

\section{Nonreciprocal systems}
\label{sec:NRS}

\begin{figure}
    \centering
    \includegraphics[width=1.0\linewidth]{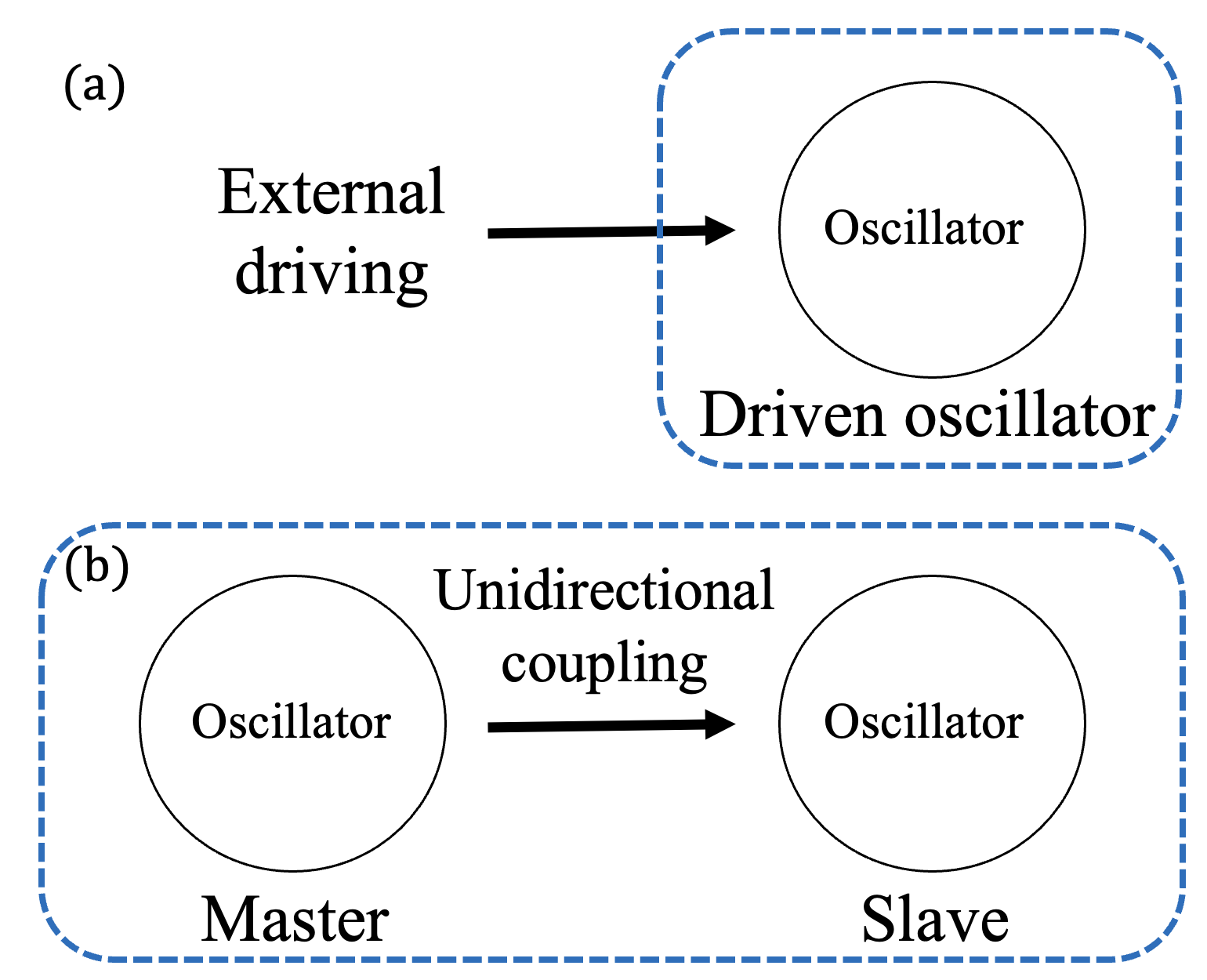}
    \caption{Schematic figures of (a) an oscillator driven by an external driving force and (b) coupled oscillators with unidirectional coupling. The dashed boxes represent the systems under consideration.}
    \label{fig:uni}
\end{figure}

{\it Inherent source in nonreciprocal systems} --- To provide an example of a nonreciprocal system, we first consider an oscillator driven by an external force. If the external force is generated by another oscillator, the overall system can be considered as master and slave oscillators with unidirectional coupling (see Fig.~\ref{fig:uni}). The master oscillator affects the slave oscillator, but the former is not affected by the latter. As a result, such nonreciprocal systems contain a kind of inherent source, which is the master oscillator. Mathematically, nonreciprocal models without gain and loss can be transformed to reciprocal models with gain and loss by similarity transformations with proper transform matrices.

{\it Nonreciprocal Hamiltonians} --- While Hermitian Hamiltonians have real eigenenergies and orthogonal eigenstates, non-Hermitian Hamiltonians including nonreciprocal models can have real or complex eigenenergies and orthogonal or non-orthogonal eigenstates. Non-Hermitian Hamiltonians satisfy the condition
\begin{equation}
    H \ne H^{\dagger}.
\end{equation}
If the Hamiltonians have real eigenenergies, then they satisfy the pseudo-Hermiticity condition \cite{Mos02a, Mos02b}
\begin{equation}
\label{eq:sh}
    \eta H \eta^{-1} = H^{\dagger},
\end{equation}
where $\eta$ is a proper transformation matrix. This is the necessary condition to have real eigenenergies, not sufficient condition. On the other hand, the condition for normal Hamiltonians with orthogonal eigenstates is
\begin{equation}
\label{eq:nn}
    H H^{\dagger} = H^{\dagger} H.
\end{equation}
Hermitian Hamiltonians have real eigenenergies and orthogonal eigenstates because they satisfy the two conditions in Eqs.~(\ref{eq:sh}) and (\ref{eq:nn}). Since the conditions for real eigenenergies and orthogonal eigenstates are independent, non-Hermitian Hamiltonians can have (i) complex eigenenergies and orthogonal eigenstates, (ii) real eigenenergies and non-orthogonal eigenstates, and (iii) complex eigenenergies and non-orthogonal eigenstates. In this work, we focus on the first two cases using one-dimensional (1D) chain models with unidirectional coupling, representing extreme cases of nonreciprocity under PBC and OBC, respectively. 

\begin{figure*}
    \centering
    \includegraphics[width=\figsizeone\linewidth]{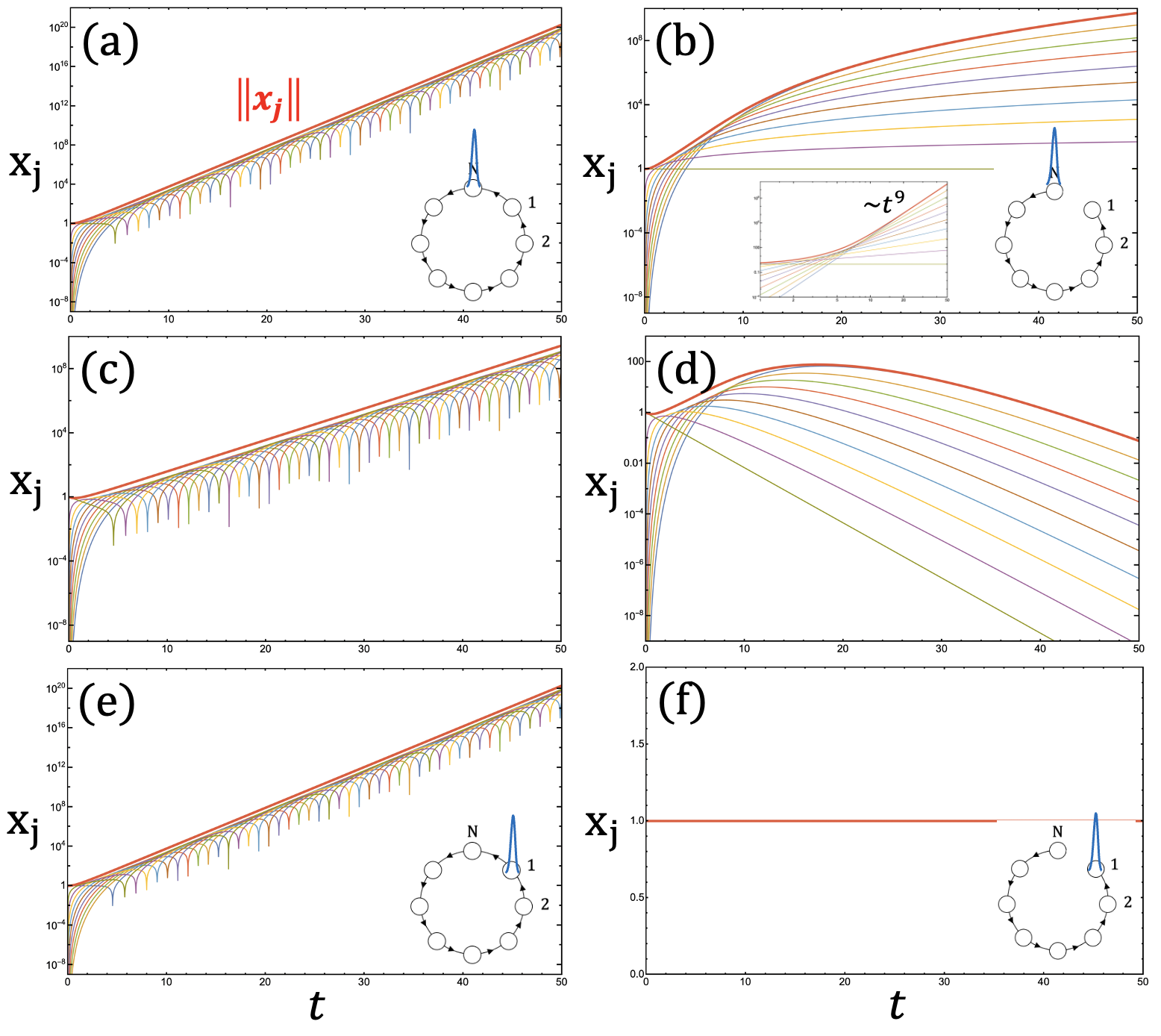}
    \caption{Time evolutions of the amplitudes of the elements $x_j (t)$ of $\left|\psi(t)\right>$ when the initial states are localized on the tenth site under (a) PBC and (b) OBC. Solid red lines represent the Euclidean norms of the states. The inset is a log-log scaled plot, $\left| x_1 (t) \right| \propto t^9$. (c,d) The same conditions with additional net loss. (e,f) The same conditions with different initial states localized on the first site. In (f), $x_j = 0 ~ (j=2,3,\dots,10)$.}
    \label{fig:right}
\end{figure*}

{\it 1D chain models with unidirectional coupling under PBC and OBC} --- We can expect that the initial states of nonreciprocal systems can be amplified as time goes by since the systems have a kind of inherent source. To demonstrate such amplifications in nonreciprocal systems, we consider simple 1D chain models with unidirectional coupling under PBC and OBC. The $N \times N$ Hamiltonian is given by
\begin{eqnarray}
\label{eq:ham}
    H=
    \begin{pmatrix}
        -i \gamma & t_l & 0 & \cdots & 0 & 0 & \beta t_r \\
        t_r & -i \gamma & t_l & \cdots & 0 & 0 & 0 \\
        0 & t_r & -i \gamma & \cdots & 0 & 0 & 0 \\
        \vdots & \vdots & \vdots & \ddots & \vdots & \vdots & \vdots \\
        0 & 0 & 0 & \cdots & -i \gamma & t_l & 0 \\
        0 & 0 & 0 & \cdots & t_r & -i \gamma & t_l \\
        \beta t_l & 0 & 0 & \cdots & 0 & t_r & -i \gamma \\
\end{pmatrix},
\end{eqnarray}
where $t_l = 1$, $t_r = 0$, $\gamma$ is net loss, and $\beta$ represents the boundary condition. The PBC is when $\beta = 1$ [see the inset in Fig.~\ref{fig:right}(a)] and the OBC is when $\beta = 0$ [see the inset in Fig.~\ref{fig:right}(b)]. Hamiltonians under PBC have complex eigenenergies and orthogonal eigenstates, while Hamiltonians under OBC have real eigenenergies and non-orthogonal eigenstates.

\section{Dynamics in nonreciprocal systems}

\subsection{Time evolution}

We consider quantum and wave systems described by the state $\left|\psi\right>$. The system dynamics is determined by the time-dependent Schr{\"o}dinger equation 
\begin{equation}
    i \hbar \frac{d}{dt} \left| \psi(t) \right> = H \left| \psi(t) \right>,
\end{equation}
where time is scaled such that $\hbar = 1$ throughout this paper. The time evolution of the state $\left|\psi(t)\right>$, which is a superposition of eigenstates or associated eigenstates in the case of defective Hamiltonians, can be described by a differential equation given by
\begin{equation}
\label{eq:diff}
    \dot{\left|\psi(t)\right>} = -i H \left|\psi(t)\right>,
\end{equation}
where the state and its derivative are
\begin{eqnarray}
\label{eq:Rstate}
    \left|\psi(t)\right> =
    \begin{pmatrix}
        x_1(t) \\
        x_2(t) \\
        \vdots \\
        x_j(t) \\
        \vdots \\
        x_N(t) \\
    \end{pmatrix}, ~
    \dot{\left|\psi(t)\right>} =
    \begin{pmatrix}
        \dot{x_1(t)} \\
        \dot{x_2(t)} \\
        \vdots \\
        \dot{x_j(t)} \\
        \vdots \\
        \dot{x_N(t)} \\
    \end{pmatrix}.
\end{eqnarray}
We assume the initial state is localized on the $N$th site [see the insets in Fig.~\ref{fig:right} (a) and (b)] and then solve the differential equation, Eq.~(\ref{eq:diff}), for $N=10$.

First, we consider the Hamiltonian without net loss, $\gamma = 0$. The Hamiltonians under PBC, i.e., Eq.~(\ref{eq:ham}) with $\beta=1$, have complex eigenenergies and orthogonal eigenstates. If an initial state is localized on the tenth site, the Euclidean norm of the state defined as
\begin{equation}
    ||x|| = \sqrt{\sum_{j=1}^{N} \left|{x_j}\right|^2}
\end{equation}
increases exponentially with time because the largest imaginary parts of the complex eigenenergies of $H$ are positive [Fig.~\ref{fig:right}(a)]. The exponents are directly related to the imaginary part of the complex eigenenergy with the largest imaginary part, $\mathrm{Max}[\mathrm{Im}(\lambda)]$, as follows:
\begin{equation}
    ||x(t)|| = e^{\mathrm{Max}[\mathrm{Im}(\lambda)] t}.
\end{equation}
On the other hand, the Hamiltonians under OBC, Eq.~(\ref{eq:ham}) with $\beta=0$, have real eigenenergies and non-orthogonal eigenstates. If an initial state is localized on the tenth site, the Euclidean norm increases algebraically with time, although all eigenenergies are real [Fig.~\ref{fig:right}(b)]. Amplifications originate from the non-normality of the systems, and as a result, the source effects of nonreciprocal systems result in the amplifications being associated with complex eigenenergies or non-orthogonal eigenstates.

To elucidate the amplification due to non-normality, we consider a simple $2 \times 2$ nonreciprocal effective Hamiltonian, 
\begin{eqnarray}
\label{eq:2by2ham}
    H=
    \begin{pmatrix}
        0 & 1 \\
        0 & 0 \\
\end{pmatrix}.
\end{eqnarray}
The time evolution of the state $\left|\psi(t)\right>$ according to Eq.~(\ref{eq:diff}) is given by
\begin{eqnarray}
    \left|\psi(t)\right> &=& e^{-i H t} \left|\psi(0)\right> \\
    \begin{pmatrix}
        x_1(t) \\
        x_2(t) \\
    \end{pmatrix} &=& 
     \begin{pmatrix}
        1 & - i t \\
        0 & 1 \\
    \end{pmatrix}
    \begin{pmatrix}
        x_1(0) \\
        x_2(0) \\
    \end{pmatrix}   
\end{eqnarray}
from the matrix exponential $e^{\mathbf{M}} = \sum_{k=0}^{\infty}{\mathbf{M}}^{k}/{k!}$. While $x_2 (t)$ does not change from the initial value $x_2 (0)$, the $x_1 (t)$ term changes with time by $-i t x_2 (0)$. The $x_1 (t)$ and $x_2 (t)$ terms are slave and master elements, respectively. Considering an $N \times N$ effective Hamiltonian, $\left|x_j (t)\right| \propto t^{N-j} x_N (0)$ [see the inset in Fig.~\ref{fig:right}(b)]. The larger the system size, the higher the exponent and the greater the amplification.

Next, we consider a Hamiltonian with net loss, $\gamma = 0.5$. In the case of PBC, the amplifying rates decrease since the largest imaginary values of the complex eigenenergies become smaller due to the net loss [Fig.~\ref{fig:right}(c)]. In the case of OBC, the amplitudes temporarily increase due to non-normality but finally decrease with time after a transient since the imaginary parts of all eigenenergies are $-0.5$. Finally, transient amplitude growth originates from the non-orthogonality of the eigenstates, but the long-time exponential decay or amplification is governed by the nonzero imaginary parts of the complex eigenenergies. We note that the long-time dynamics is also governed by non-normality if all eigenenergies are real since there is no long-time exponential decay or amplification resulting from the complex eigenenergies [see Fig.~\ref{fig:right}(b)].

To reconsider the amplifications, we introduce a simple $2 \times 2$ nonreciprocal effective Hamiltonian with net loss,
\begin{eqnarray}
\label{eq:2by2ham}
    H=
    \begin{pmatrix}
        - i \gamma & 1 \\
        0 & - i \gamma \\
\end{pmatrix}.
\end{eqnarray}
The time evolution of the state $\left|\psi(t)\right>$ is given by
\begin{eqnarray}
    \begin{pmatrix}
        x_1(t) \\
        x_2(t) \\
    \end{pmatrix} &=& 
     \begin{pmatrix}
        e^{-\gamma t} & - i t e^{-\gamma t} \\
        0 & e^{-\gamma t} \\
    \end{pmatrix}
    \begin{pmatrix}
        x_1(0) \\
        x_2(0) \\
    \end{pmatrix}.   
\end{eqnarray}
While $x_2 (t)$ decays exponentially with exponent $-\gamma t$ from the initial value $x_2 (0)$, $x_1 (t)$ decays exponentially with exponent $-\gamma t$ with transient growth by the term $-i t e^{-\gamma t} x_2 (0)$.

The transient growth depends on the initial states, while the long-time exponential behavior is irrespective of the initial states. This is natural since the transient growth and long-time exponential behavior are related to the eigenstates and eigenenergies, respectively. If the initial states are localized on the first site, the results from the Hamiltonian under PBC are not different from those when the initial states are localized on the tenth site [see Fig.~\ref{fig:right}(e)]. Under OBC, however, if the initial states are localized on the first site, then the initial states do not change with time since there is no influence from the states localized on the first site to other sites. This coincides with the fact that the initial states are the right eigenstates of the Hamiltonian and the corresponding eigenenergies are real. If there is net loss, the initial states localized on the first site exponentially decay without abnormal transient behavior. As a result, the initial states leading to the maximum transient growth correspond to left eigenstates (states localized on the tenth site), while right eigenstates as the initial states cannot achieve transient growth but always follow the imaginary part of the associated complex eigenenergy. For example, if we consider initial states leading to the maximum transient growth in a $2\times2$ passive PT-symmetric Hamiltonian with net loss, while the initial right eigenstates exponentially decay, the initial left eigenstates temporarily amplify and finally decay at the same rate (see Appendix).  

\begin{figure*}
    \centering
    \includegraphics[width=\figsizeone\linewidth]{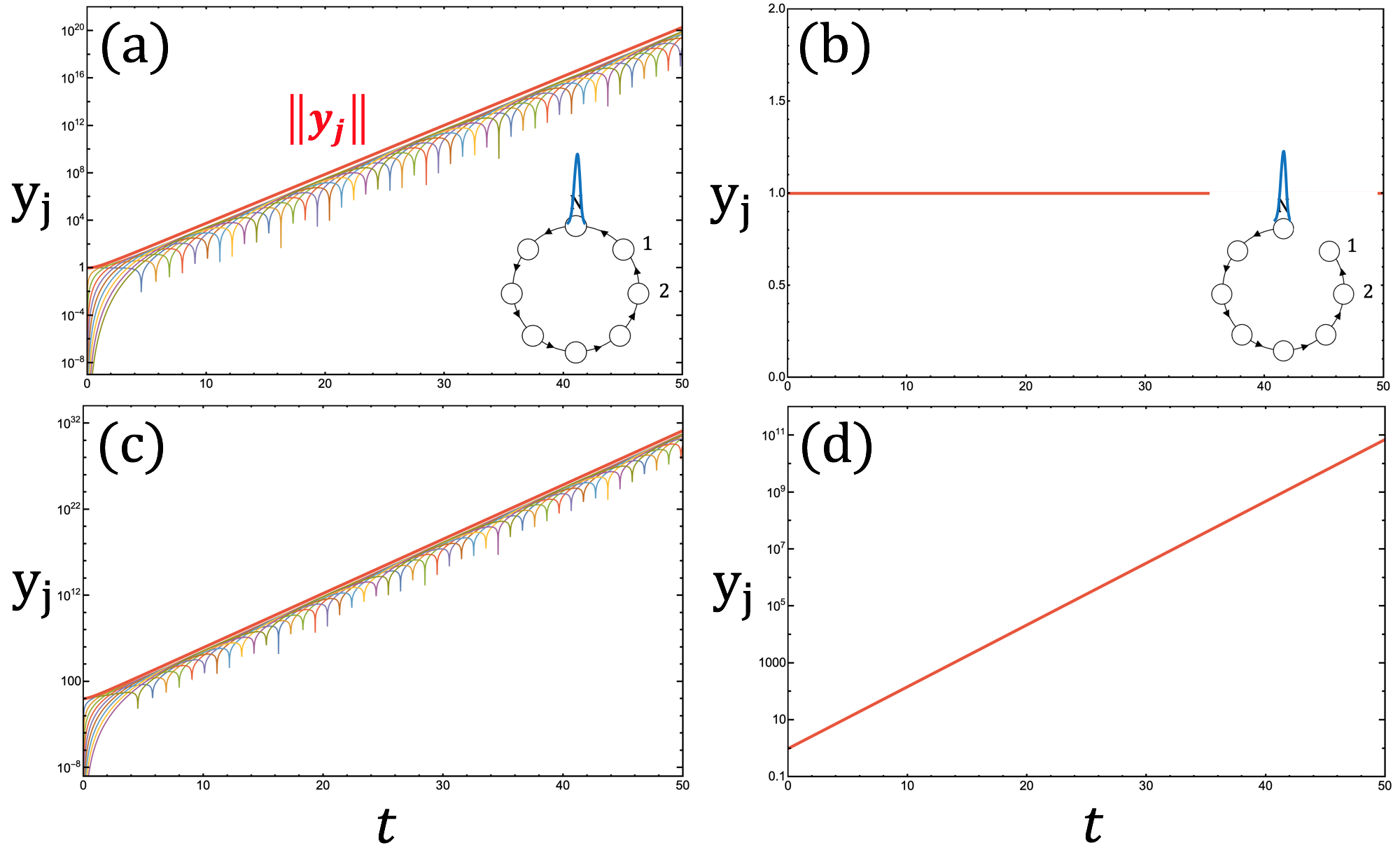}
    \caption{Time evolutions of the amplitudes of the elements $y_j (t)$ of $\left|\phi\right>$ when the initial states are localized on the tenth site under (a) PBC and (b) OBC. Solid red lines represent the Euclidean norms of the states. (c,d) The same conditions with additional net loss. In (b) and (d), $y_j = 0 ~ (j=1,2,\dots,9)$.}
    \label{fig:left}
\end{figure*}

\subsection{Time evolution in terms of left eigenstates}

Many physical measures such as the amplitudes of oscillators \cite{Ryu17, Ryu19} and waves \cite{Wie11, Wie18} described by coupled differential equations analogous to Eq.~(\ref{eq:diff}) can be well represented by Eq.~(\ref{eq:Rstate}) associated with right eigenstates. For instance, linearized Jacobian matrix $J$ in a coupled nonlinear oscillator model is mathematically equivalent to $-iH$ in Eq.~(\ref{eq:diff}), while their physical meanings are different. However, the biorthogonality associated with the left as well as the right eigenstates is required to understand non-Hermitian quantum systems \cite{Lee20}. We consider the time-dependent Schr{\"o}dinger equation
\begin{equation}
    i \hbar \frac{d}{dt} \left| \phi(t) \right> = H^\dagger \left| \phi(t) \right>,
\end{equation}
where $\left|\phi(t)\right>$ is a superposition of the right eigenstates of $H^{\dagger}$, which are the same as the left eigenstates of $H$, while $\left|\psi(t)\right>$ is a superposition of the right eigenstates of $H$. The time evolution of the state $\left| \phi(t) \right>$ can be described by a differential equation given by
\begin{equation}
    \dot{\left|\phi(t)\right>} = -i H^\dagger \left|\phi(t)\right>,
\end{equation}
where the state and its derivative are
\begin{eqnarray}
\label{eq:Lstate}
    \left|\phi(t)\right> =
    \begin{pmatrix}
        y_1(t) \\
        y_2(t) \\
        \vdots \\
        y_j(t) \\
        \vdots \\
        y_N(t) \\
    \end{pmatrix}, ~
    \dot{\left|\phi(t)\right>} =
    \begin{pmatrix}
        \dot{y_1(t)} \\
        \dot{y_2(t)} \\
        \vdots \\
        \dot{y_j(t)} \\
        \vdots \\
        \dot{y_N(t)} \\
    \end{pmatrix}.
\end{eqnarray}

The amplitudes of the state $\left|\phi (t)\right>$ evolving from the initial states localized on the tenth (first) site are the same as the amplitudes of the state $\left|\psi (t)\right>$ evolving from the initial states localized on the first (tenth) site, irrespective of the boundary conditions [see Fig.~\ref{fig:left} (a) and (b)]. If we add net loss (gain) to the Hamiltonian $H$ ($H^\dagger$), the amplification rate becomes larger since the corresponding eigenenergy of the left eigenstates is a complex conjugate of that of the right eigenstates. The Euclidean norm of the state $\left|\phi (t)\right>$ increases with time, contrary to the case of the state $\left|\psi (t)\right>$ [see Fig.~\ref{fig:left} (c) and (d)].

\begin{figure*}
    \centering
    \includegraphics[width=\figsizeone\linewidth]{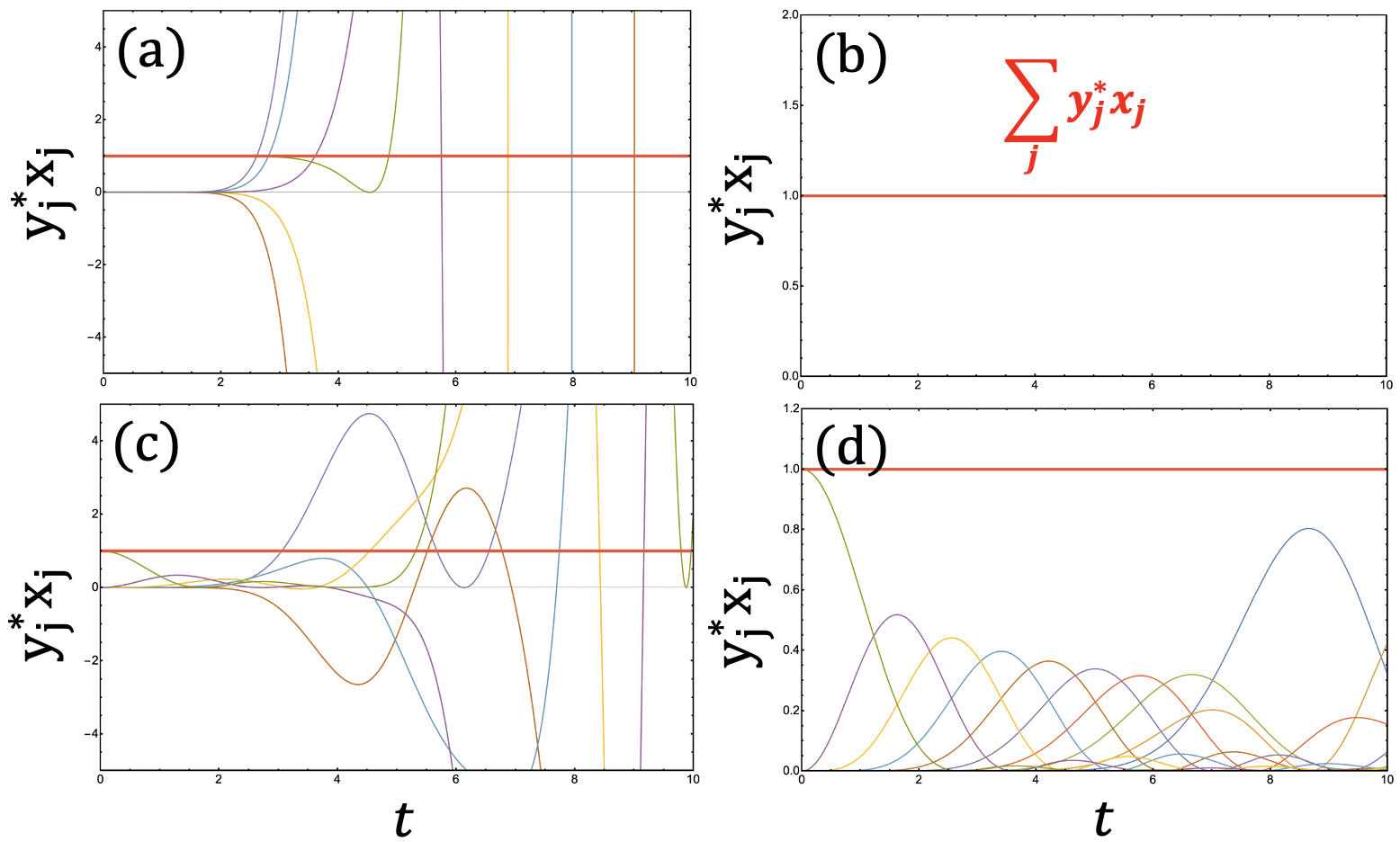}
    \caption{Time evolutions of the amplitudes of the elements $y_{j}^{*} x_j$ when the initial states are localized on the tenth site under (a) PBC and (b) OBC. Solid red lines represent the biorthogonal norms of the states. In (b), $y_j^{*} x_j = 0 ~ (j=1,2,\dots,9)$ since $y_j  = 0 ~ (j=1,2,\dots,9)$. (c,d) The same conditions with partially directed coupling, $t_l = 1.0$ and $t_r = 0.5$, instead of unidirectional coupling.}
    \label{fig:bi}
\end{figure*}
  
\subsection{Time evolution in terms of biorthogonality}

The Euclidean norms of $\left|\psi (t)\right>$ associated with the right eigenstates and of $\left|\phi (t)\right>$ associated with the left eigenstates are no longer equal to $1$ in nonreciprocal systems since they have complex eigenenergies or non-orthogonal eigenstates. However, biorthogonal norms always satisfy 
\begin{equation}
\label{bi_norm}
    \sum_{j}{y_{j}^{*} x_j} = 1,
\end{equation}
because of biorthogonality in non-Hermitian systems. Additionally, $y_{j}^{*} x_j$ is real. The $y_{j}^{*} x_j$ term can be larger than $1$ and negative if eigenenergies are complex, while $0 < y_{j}^{*} x_j < 1$ if eigenenergies are real (see Fig.~\ref{fig:bi}). In the case of OBC with unidirectional coupling $t_l = 1.0$ and $t_r = 0.0$ with the initial states localized on the tenth site, $y_{j}^{*} x_j = 0 ~(j=1,2,\dots,9)$ and $y_{10}^{*} x_{10} = 1$ since $y_j = 0 ~(j=1,2,\dots,9)$ [see Fig.~\ref{fig:bi}(b)].

\begin{figure*}
    \centering
    \includegraphics[width=1.0\linewidth]{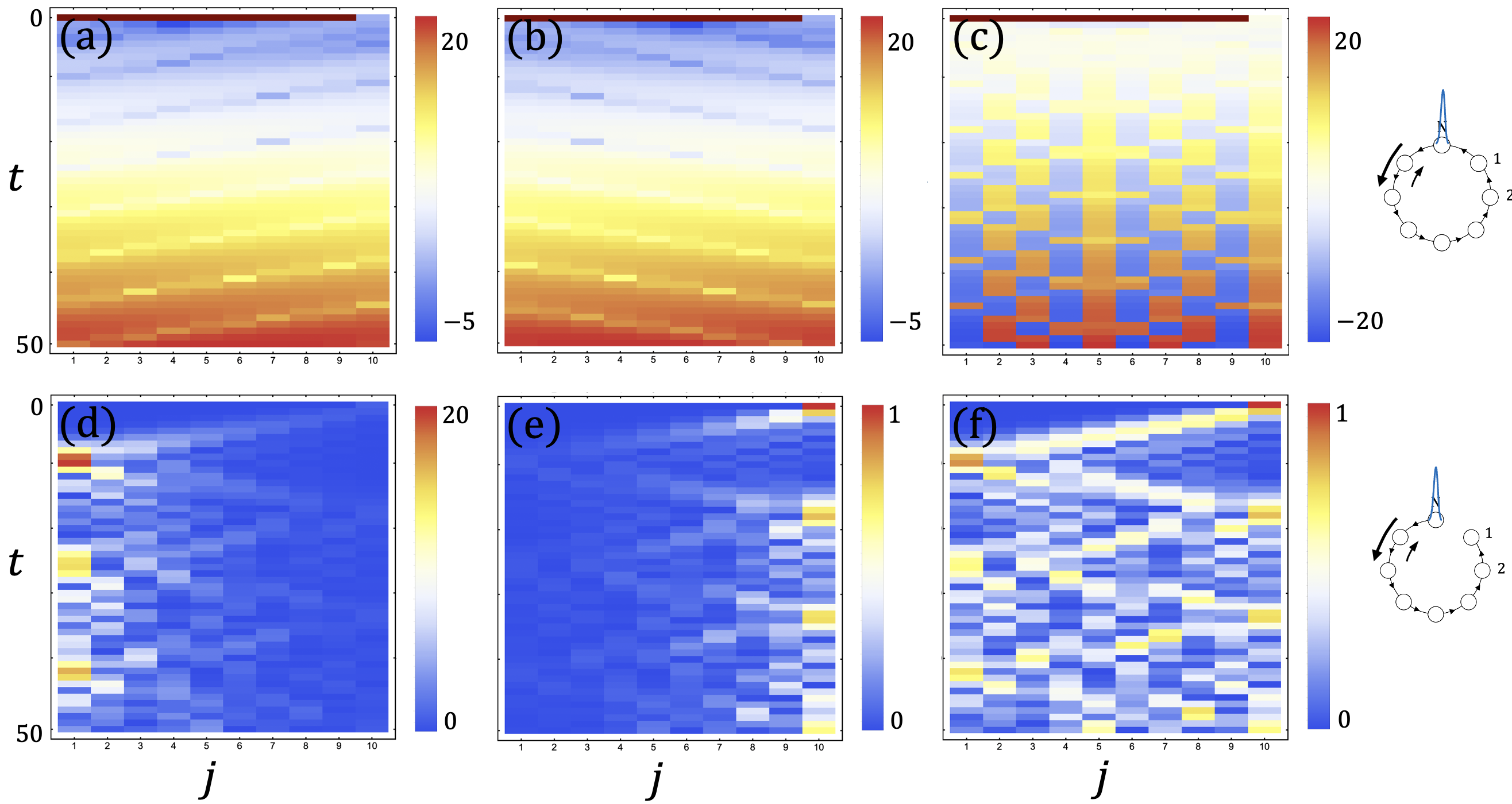}
    \caption{Time evolutions of the amplitudes of the elements (a) $\log \sqrt{x_j^{*} x_j} $, (b) $\log \sqrt{y_j^{*} y_j} $, and (c) $\mathrm{sgn}(y_{j}^{*} x_j)\log \sqrt{|y_{j}^{*} x_j|} $ when the initial states are localized on the tenth site under PBC with partially directed coupling, $t_l = 1.0$ and $t_r = 0.5$. The brown colors when $t=0$ represent zero initial amplitudes. Time evolutions of the amplitudes of the elements (d) $\sqrt{x_j^{*} x_j}$, (e) $\sqrt{y_j^{*} y_j}$, and (f) $\sqrt{y_{j}^{*} x_j}$ when the initial states are localized on the tenth site under OBC with the same partially directed coupling.}
    \label{fig:directional}
\end{figure*}

\subsection{Time evolution with partially directed coupling}

We now consider partially directed coupling, $t_l = 1$ and $t_r = 0.5$. In the case of PBC, the amplitudes $\sqrt{x_j^{*} x_j} $ of the elements of $\left|\psi (t)\right>$ rotate in the counterclockwise direction, while the amplitudes $\sqrt{y_j^{*} y_j} $ of the elements of $\left|\phi (t)\right>$ rotate in the clockwise direction. Their amplitudes increase with time since the largest imaginary value of complex eigenenergies is positive (see Fig.~\ref{fig:directional}). The amplitude $\sqrt{y_j^{*} x_j}$ of the biorthogonal elements combines the behaviors of $x_j$ and $y_j$. They rotate in both counterclockwise and clockwise directions, and their spatial inhomogeneity also increases with time. It is noted that biorthogonal norms always satisfy Eq.~(\ref{bi_norm}). In the case of OBC, $\sqrt{x_j^{*} x_j} $ rotates in the counterclockwise direction and bounces at the end of the chain (the first site). The states repeat this as time goes by [see Fig.~\ref{fig:directional}(d)]. Amplification of the states in the partially directed coupling case is saturated, while the states in the unidirectional coupling case increase algebraically without saturation [see Fig.~\ref{fig:right}(b)]. Also, $\sqrt{y_j^{*} y_j} $ rotates in the clockwise direction and bounces at the opposite end of the chain (the tenth site), and the states repeat this as time goes by. Hence $\sqrt{y_j^{*} x_j}$ combines the behaviors of $x_j$ and $y_j$ but all amplitudes of $y_j^{*} x_j$ satisfy $0 < y_j^{*} x_j < 1$ since all eigenenergies are real [see Fig.~\ref{fig:bi}(d) and Fig.~\ref{fig:directional}(f)]. We note that these results are in line with recent experimental results with nonreciprocal robotic metamaterials \cite{Bra19}.

\section{Summary}

We have studied dynamics in nonreciprocal systems having an inherent source due to directed coupling. We showed that amplifications with time without net loss can appear due to complex eigenvalues and non-orthogonal eigenvectors, while transient amplifications with net loss can appear depending on the initial states. The abnormal time evolutions were elaborated using different measures defined by the right eigenstates, left eigenstates, and biorthogonal states. We expect more interesting phenomena to emerge if nonlinearity and inhomogeneous external pumping are applied to our linear models. 

\section*{acknowledgments}
The author thanks J.-H. Han, B. H. Kim, K.-M. Kim, S. Park, H. C. Park, and C.-H. Yi for helpful discussions. 
Financial support is acknowledged from the Institute for Basic Science in the Republic of Korea through the project IBS-R024-D1.

\section{Appendix}

\subsection{Transient growth in a $2 \times 2$ passive PT-symmetric Hamiltonian in terms of initial states}
We consider the initial states leading to the maximum transient growth in a $2\times2$ passive PT-symmetric Hamiltonian with a net loss of $\gamma_0$,
\begin{eqnarray}
        H=
    \begin{pmatrix}
        i \gamma - i \gamma_0 & c  \\
        c & -i \gamma- i \gamma_0  \\
    \end{pmatrix},  
\end{eqnarray}
where $\gamma = c = 1$ and the eigenenergies are $-i \gamma_0$ at the exceptional point. The right and left eigenstates are $(i,1)^T$ and $(-i,1)^T$, respectively. The initial right eigenstates exponentially decay, i.e., $e^{-\gamma_0 t}$, but the initial left eigenstates temporarily amplify and finally decay at the same rate (not shown).


\newpage

\begin{thebibliography}{150}

\bibitem{Xu19} H. Xu, L. Jiang, A. A. Clerk, and J. G. E. Harris, Nonreciprocal control and cooling of phonon modes in an optomechanical system, Nature {\bf 568}, 65–69 (2019).
\bibitem{Pot04} R. J. Potton, Reciprocity in optics. Rep. Prog. Phys. {\bf 67}, 717–754 (2004).
\bibitem{Est14} N. A. Estep, D. L. Sounas, J. Soric, and A. Alù, Magnetic-free non-reciprocity and isolation based on parametrically modulated coupled-resonator loops, Nat. Phys. {\bf 10}, 923–927 (2014).
\bibitem{Rue16} F. Ruesink, M. A. Miri, A. Alù, and E. Verhagen, Nonreciprocity and magnetic-free isolation based on optomechanical interactions, Nat. Commun. {\bf 7}, 13662 (2016).
\bibitem{Kod13} T. Kodera, D. L. Sounas, and C. Caloz, Magnetless nonreciprocal metamaterial (mnm) technology: application to microwave components, IEEE Trans.
Microw. Theory Techn. {\bf 61}, 1030–1042 (2013).

\bibitem{Fle14} R. Fleury, D. L. Sounas, C. F. Sieck, M. R. Haberman, and A. Alù, Sound isolation and giant linear nonreciprocity in a compact acoustic circulator,
Science {\bf 343}, 516–519 (2014).
\bibitem{Wan18} Y. Wang, B. Yousefzadeh, H. Chen, H. Nassar, G. Huang, and C. Daraio, Observation of nonreciprocal wave propagation in a dynamic phononic lattice. Phys. Rev. Lett. {\bf 121}, 194301 (2018).
\bibitem{Sou18} D. L. Sounas, J. Soric, and A. Alù, Broadband passive isolators based on coupled nonlinear resonances. Nat. Electron {\bf 1}, 113–119 (2018).
\bibitem{Pea16} V. Peano, M. Houde, F. Marquardt, and A. A. Clerk, Topological quantum fluctuations and traveling wave amplifiers, Phys. Rev. X {\bf 6}, 041026 (2016).
\bibitem{Lau18} H. K. Lau, and A. A. Clerk, Fundamental limits and nonreciprocal approaches in non-hermitian quantum sensing. Nat. Commun. {\bf 9}, 4320 (2018).
\bibitem{Nas15} L. M. Nash, D. Kleckner, A. Read, V. Vitelli, A. M. Turner, and W. T. M. Irvine, Topological mechanics of gyroscopic metamaterials, Proc. Natl Acad. Sci. USA{\bf 112}, 14495–14500 (2015).
\bibitem{Cou17} C. Coulais, D. Sounas, and A. Alù, Static non-reciprocity in mechanical metamaterials, Nature {\bf 542}, 461–464 (2017).
\bibitem{Bra19} M. Brandenbourger, X. Locsin, E. Lerner, and C. Coulais, nonreciprocal robotic metamaterials, Nat. Commun. {\bf 10}, 4608 (2019).


\bibitem{Hat96} N. Hatano and D. R. Nelson, Localization transitions in non-Hermitian quantum mechanics, Phys. Rev. Lett. {\bf 77}, 570 (1996).
\bibitem{Hat97} N. Hatano and D. R. Nelson, Vortex pinning and non-Hermitian quantum mechanics, Phys. Rev. B {\bf 56}, 8651 (1997). 
\bibitem{Yao18a} S. Yao and Z. Wang, Edge states and topological invariants of non-Hermitian systems, Phys. Rev. Lett. {\bf 121}, 086803 (2018)
\bibitem{Yao18b} S. Yao, F. Song, and Z. Wang, Non-Hermitian Chern Bands, Phys. Rev. Lett. {\bf 121}, 136802 (2018).
\bibitem{Yan19} Z. Yang and J. Hu, Non-Hermitian Hopf-link exceptional line semimetals, Phys. Rev. B {\bf 99}, 081102(R) (2019).
\bibitem{Jin19} L. Jin and Z. Song, Bulk-boundary correspondence in a non-Hermitian system in one dimension with chiral inversion symmetry, Phys. Rev. B {\bf 99}, 081103(R) (2019).
\bibitem{Wan19} H. Wang, J. Ruan, and H. Zhang, Non-Hermitian nodal-line semimetals with an anomalous bulk-boundary correspondence, Phys. Rev. B {\bf 99}, 075130 (2019).
\bibitem{Lee19} C. H. Lee and R. Thomale, Anatomy of skin modes and topology in non-Hermitian systems, Phys. Rev. B {\bf 99}, 201103(R) (2019).
\bibitem{Gha19} A. Ghatak and T. Das, New topological invariants in non-Hermitian systems, J. Phys.: Condens. Matter {\bf 31}, 263001 (2019).
\bibitem{Gha21} H. Ghaemi-Dizicheh and H. Schomerus, Compatibility of transport effects in non-Hermitian nonreciprocal systems, Phys. Rev. A {\bf 104}, 023515 (2021).

\bibitem{Tre05} L. N. Trefethen and M. Embree, {\it Spectra and Pseudospectra} (Princeton University Press, Princeton, NJ, 2005).

\bibitem{Tre93} L. N. Trefethen, A. E. Trefethen, S. C. Reddy, and T. A. Driscoll, Hydrodynamic stability without eigenvalues, Science {\bf 261}, 578 (1993).
\bibitem{Tre97} L. N. Trefethen, Pseudospectra of linear operators, SIAM Rev.
{\bf 39}, 383 (1997).

\bibitem{Neu02} M. G. Neubert, H. Caswell, and J. D. Murray, Transient dynamics and pattern formation: reactivity is necessary for Turing instabilities, Math. Biosci. {\bf 175}, 1 (2002).
\bibitem{Rid11} L. Ridolfi, C. Camporeale, P. D’Odorico, and F. Laio, Transient growth induces unexpected deterministic spatial patterns in the Turing process, Europhys. Lett. {\bf 95}, 18003 (2011).

\bibitem{Bia17} T. Biancalani, F. Jafarpour, and N. Goldenfeld, Giant Amplification of Noise in Fluctuation-Induced Pattern Formation, Phys. Rev. Lett.
{\bf 118}, 018101 (2017).

\bibitem{Mak14} K. G. Makris, L. Ge, and H. E. Türeci, Anomalous Transient Amplification of Waves in Non-normal Photonic Media, Phys. Rev. X {\bf 4}, 041044 (2014). 
\bibitem{Mak21} K. G. Makris, Transient growth and dissipative exceptional points, Phys. Rev. E {\bf 104} 054218 (2021).

\bibitem{Asl18a} M. Asllani and T. Carletti, Topological resilience in non-normal networked systems, Phys. Rev. E {\bf 97}, 042302 (2018).
\bibitem{Asl18b} M. Asllani, R. Lambiotte, and T. Carletti, Structure and dynamical behavior of non-normal networks, Sci. Adv. {\bf 4} eaau9403 (2018).
\bibitem{Bag20} G. Baggio, V. Rutten, G. Hennequin, and S. Zampieri, Efficient communication over complex dynamical networks: The role of matrix non-normality, Sci. Adv. {\bf 6} eaba2282 (2020).
\bibitem{Mou21} R. Muolo, T. Carletti, J. P. Gleeson, and M. Asllani, Synchronization Dynamics in Non-Normal Networks: The Trade-Off for Optimality, Entropy {\bf 23}, 36 (2021).
\bibitem{Dua22} C. Duan, T. Nishikawa, D. Eroglu, and A. E. Motter, Network structural origin of instabilities in large complex systems, Sci. Adv. {\bf 8}, eabm8310(2022).

\bibitem{Mos02a} A. Mostafazadeh, Pseudo-Hermiticity versus PT symmetry: The necessary condition for the reality of the spectrum of a non-Hermitian Hamiltonian, J. Math. Phys. {\bf 43}, 205–214 (2002)
\bibitem{Mos02b} A. Mostafazadeh, Pseudo-Hermiticity for a class of nondiagonalizable Hamiltonians, J. Math. Phys. {\bf 43}, 6343–6352 (2002).

\bibitem{Ryu17} J.-W. Ryu, J.-H. Kim, W.-S. Son, D.-U. Hwang, Amplitude death in a ring of nonidentical nonlinear oscillators with unidirectional coupling, Chaos 27, 083119 (2017).
\bibitem{Ryu19} J.-W. Ryu, W.-S. Son, and D.-U. Hwang, Oscillation death in coupled counter-rotating identical nonlinear oscillators, Phys. Rev. {\bf E} 100, 022209 (2019).
\bibitem{Wie11} J. Wiersig, Structure of whispering-gallery modes in optical microdisks perturbed by nanoparticles, Phys. Rev. A {\bf 84}, 063828 (2011).
\bibitem{Wie18} J. Wiersig, Role of nonorthogonality of energy eigenstates in quantum systems with localized losses, Phys. Rev. A {\bf 98}, 052105 (2018).
\bibitem{Lee20} E. Lee, H. Lee, and B.-J. Yang, Many-body approach to non-Hermitian physics in fermionic systems, Phys. Rev. B {\bf 101}, 121109(R) (2020).

\end{thebibliography}
\end{document}